\RequirePackage{fix-cm}
\documentclass[natbib,smallextended]{svjour3}       
\smartqed  
\usepackage{graphicx}
\usepackage{amssymb}
\usepackage{color}
\usepackage{isotope}
\usepackage{verbatim}

\def\aj{AJ}%
\def\apj{ApJ}%
\def\apjl{ApJ}%
\def\aap{A\&A}%
\def\mnras{MNRAS}%
\def\prc{Phys.~Rev.~C}%
\def\pasp{PASP}%
\def\nat{Nature}%
\begin{document}

\title{Models for Type Ia supernovae and related astrophysical transients
}

\titlerunning{Models for SNe~Ia}        

\author{Friedrich K.\ R{\"o}pke         \and
        Stuart A.\ Sim 
}


\institute{F.~K.~R{\"o}pke \at
  Zentrum f{\"u}r Astronomie der Universit{\"a}t Heidelberg,
  Philosophenweg 12, 69120 Heidelberg, Germany\\
  and\\
  Heidelberger Institut f{\"u}r Theoretische Studien,
  Schloss-Wolfsbrunnenweg 35, 69118 Heidelberg, Germany\\
              \email{friedrich.roepke@h-its.org}           
           \and
           S. A. Sim \at
              School of Mathematics and Physics, Queen's University Belfast, Belfast BT7 1NN, UK
}

\date{Received: date / Accepted: date}

\maketitle

\begin{abstract}
We give an overview of recent efforts to model Type Ia supernovae and
related astrophysical transients resulting from thermonuclear
explosions in white dwarfs. In particular we point out the challenges
resulting from the multi-physics multi-scale nature of the problem and
discuss possible numerical approaches to meet them in hydrodynamical
explosion simulations and radiative transfer modeling. We give
examples of how these methods are applied to several explosion
scenarios that have been proposed to explain distinct subsets or, in
some cases, the majority of the observed events. In case we comment on
some of the successes and shortcoming of these scenarios and highlight
important outstanding issues.
\end{abstract}

\section{Introduction}
\label{intro}

The theoretical description of Type Ia supernovae and related
astrophysical transients as thermonuclear explosions of white dwarfs
stars has seen rapid development over the past decade.
Multidimensional hydrodynamical simulations of the explosion phase were
conducted, and the results could be directly used as input for
radiative transfer simulations that derive synthetic observables from
such models in a consistent way. This allowed to connect modern
supernova theory directly to astronomical observations and facilitated
a way to validate modeling assumptions by comparison with astronomical
data.

The result of theoretical efforts is a consistent theoretical modeling
pipeline for thermonuclear explosions in white dwarf stars. It starts
out from a model of the progenitor and extends over multidimensional
hydrodynamical simulations of the explosion phase. Nucleosynthesis
processes in it are usually determined in a post-processing step. This
gives a multidimensional picture of the structure (in particular the
density, the velocity and the chemical composition) of the ejecta
cloud, that serves as an input to radiative transfer
calculations. These, in turn, allow to derive synthetic observables.

In the two parts of this article, we discuss two main ingredients to
this modeling pipeline: hydrodynamics simulations of the explosion
phase together with nucleosynthesis calculations, and the radiative
transfer in the ejecta.

\section{Explosion modeling}
\subsection{Ansatz and scale challenges}
\label{sect:hydro_challenges}

The progenitor star of a thermonuclear supernova event is
a macroscopic object; densities are high and the typical spatial
scales of interest are large. Therefore, the modeling ansatz is based
on the equations of fluid dynamics, specifically the Euler equations
describing ideal fluids. Viscosity effects are sub-dominant on the
scales considered in the model and numerical viscosity in any case
outweighs physical effects.

As nuclear burning powers the supernova explosion, reactions have to
be taken into account. In addition to the usual fluid dynamics
equations describing mass conservation, momentum, and energy balance,
a set of equations is necessary to capture species balance. Source
terms account for species conversion and the associated energy
release. For a complete description of thermonuclear combustion
processes, several other effects, such as heat conduction, have to be
included in the model (see, e.g., \citealp{roepke2017a} for a recent
  overview). The solution of this system of equations in
a numerical supernova simulation, however, is not straightforward
because of massive challenges arising from the extremely wide range of
relevant scales.

For a numerical treatment, the underlying system of equations is
discretized, most commonly in a finite-volume approach. The supernova
explosion takes place on the order of the dynamical time scale and
therefore time discretization usually follows an explicit scheme. Such
schemes are only conditionally stable and consequently the numerical
time step has to be restricted according to the
Courant-Friedrichs-Lewi (CFL) condition, which, in loose terms,
requires numerical time steps to be taken smaller than the sound
crossing time over a computational grid cell. The sound crossing time
over an entire WD star is on the order of a second; hence the
numerical time steps in supernova simulations stay far below a
second. The evolution time scale of the progenitor, in contrast, is
set by nuclear burning, that lasts many orders of magnitude longer
than this. Clearly, this phase is not accessible to multidimensional
hydrodynamic simulations. Depending on the propagation mode of the
resulting thermonuclear combustion wave, the ignition itself may take
centuries or happen dynamically. Thus, ignition is a marginal case
that may be addressed in the framework of hydrodynamical simulations,
at least as much as time scales are concerned. The explosion itself
proceeds in the transonic regime and is certainly accessible to such a
numerical treatment.

The time scales on which most observables form are much longer --
days, weeks, or months. Because the supernova ejecta are in homologous
expansion by then (hydrodynamical effects are frozen out) and the
radiation field is dynamically unimportant (at least to zeroth order;
see \citealp{woosley2007b} for a discussion of the effect of $^{56}$Ni
decay on the density and velocity profiles), this can be treated in a
modeling approach that is separated from the hydrodynamical
simulations of the explosion phase and uses their results only to
define the background state of the expanding ejecta (see
Sect.~\ref{sect:obs}).

The spatial scale problem in thermonuclear supernova models is no less
challenging. Due to the extreme temperature sensitivity of the
involved nuclear reactions, burning is confined to the hottest regions
and propagates in thin fronts. Typically, these have widths far below
the millimeter scale. This scale is extremely small compared with that
of the exploding white dwarfs (with radii of a few thousand
kilometers). Seen from the large global scales, it is well-justified to
approximate combustion waves as sharp discontinuities separating the
fuel from the ash material. In this discontinuity approximation, jump
conditions over the combustion front can be established according to
the laws of fluid dynamics. They distinguish between two modes of
propagation for the combustion front: subsonic \emph{deflagration} and
supersonic \emph{detonation}.

Both deflagrations and detonations are subject to multidimensional
hydrodynamic instabilities \citep[for a recent
review see][]{roepke2017a}. While for the latter case, it is generally assumed that the
effects on the overall explosion process are weak, deflagration
burning is most likely dominated -- and as a consequence significantly
boosted -- by the interaction with such instabilities.  If ignited
near the center of the white dwarf star, a deflagration becomes
turbulent. This is an implication of buoyancy instability between the
central hot and light ashes and the dense and cold unburnt fuel ahead
of the flame. As a result, in the non-linear regime of the
Landau-Rayleigh-Taylor instability, bubbles of burning material rise
towards the stellar surface (but see \citealp{hristov2017a}). The flame
front is located at their interfaces. Outside of the bubbles, cold unburnt
material sinks down towards the center of the white dwarf. This leads
to shear motions at the flame. Typical Reynolds numbers are as high as
$10^{14}$ and consequently a \emph{turbulent energy cascade} forms. At
the largest scales, kinetic energy is injected by large-scale
turbulent eddies, that subsequently decay to smaller scales
constituting the inertial range, in which kinetic energy is
transported from the large to the small scales without energy
loss. Only at the microscopic Kolmogorov scale, the turbulent energy
is finally converted to heat by viscous effects.

On a wide sub-range of that turbulent cascade, the deflagration flame
interacts with turbulent eddies (see \citealp{roepke2009a} and \citealp{roepke2017a}
for discussions of turbulent deflagrations in SNe~Ia). The effect of
this interaction depends on whether turbulence corrugates the flame
structure only on large scales, or whether it penetrates the internal
flame structure and modifies the microphysical transport in it. The
first case, which corresponds to the so-called \emph{flamelet regime}
of turbulent combustion, applies to most of the explosion
period. Here, the flame front is stretched out and wrinkled so that
its surface area is greatly enlarged.

Only at the latest times, when the star has expanded significantly and
the burning densities are low, the flame structure broadens. With the
expansion, turbulence gradually freezes out, but if the prevailing
turbulent intensities are still high, a modification of the flame
structure is expected. It has been suggested
\citep[e.g.][]{khokhlov1997a,lisewski2000b,roepke2007a,woosley2007a,schmidt2010a,poludnenko2011a}
that in this regime transitions of the flame propagation mode from
subsonic deflagration to supersonic detonation are possible. Such
\emph{deflagration-to-detonation transitions (DDTs)} are observed in
terrestrial chemical combustion, but there they are mostly associated
with obstacles or walls of the combustion vessel. The existence of
unconfined DDTs, as would be required in the astrophysical context,
remains unproven. Sufficiently strong turbulent mixing inside a broad
flame structure is proposed to lead to conditions in which a
detonation wave can form via the Zel'dovich gradient mechanism
\citep{zeldovich1970a}.

In addition to these uncertainties in the flame propagation mechanism,
the problem of the initial conditions poses a fundamental challenge to
modeling thermonuclear supernova explosions. As to now, progenitor
systems of Type Ia supernovae are not observationally
established. Although the astronomical identification of a progenitor
would help to constrain potential scenarios, it would not completely
solve all problems of initial conditions for explosion
simulations. The progenitor structure and the ignition process are not
directly accessible to observations and have to be modeled. As
discussed above, the timescales dominating the pre-ignition evolution
phases cannot easily be addressed in multidimensional simulations. The
resulting uncertainty in the initial conditions is a fundamental
obstacle to explosion modeling. The equations of hydrodynamics forming
the basis for the description of the explosion processes are
hyperbolic partial differential equations. Thus they pose initial
value problems. The choice of the initial conditions therefore has a
strong impact on the numerical solution (or even determines it). One
should thus avoid to draw conclusions from thermonuclear supernova
simulations that are dominated by an arbitrary of the initial
conditions.

\subsection{Numerical implementation}

Several approaches have been taken by different groups to meet the
challenges laid out above and perform simulations of thermonuclear
supernova explosions. An overview of modeling the combustion physics
is given in \citet{roepke2017a}. Here, we will focus on one particular
choice.

The impracticality to resolve the tiny internal structure of
combustion waves in full-star supernova explosion simulations requires
to model their propagation in parametrized approaches. The physical
structure is either artificially broadened so that it can be
represented on the computational grid
\citep{khokhlov1995a,vladimirova2006a,calder2007a}, or it is completely
ignored and the combustion front is treated as a sharp discontinuity
separating the fuel from the ashes. An appropriate technique to
achieve this (at least up to the spread in hydrodynamical quantities
introduced by the numerical Riemann solver) is the so-called level-set
scheme \citep{osher1988a,reinecke1999a}. In this front-tracking method,
the combustion wave is associated with the zero level-set of a signed
distance function $G$. Its motion is due to advection of the $G$-field
and due to burning. This is captured by an appropriate ``level-set
equation''. While the advection part can be determined from the
underlying hydrodynamics scheme, the advancement due to burning is not
consistently treated in the discontinuity approximation. It is a
parameter of the model that has to be determined externally. For
laminar deflagration flames, for instance, it can be derived from
resolved one-dimensional simulations \citep[e.g.][]{timmes1992a}. For
detonations, the Chapman-Jouguet case is a reasonable approximation at
low fuel densities. At higher densities, however, nuclear statistical
equilibrium establishes behind the detonation front and reactions are
partially endothermic. This gives rise to detonations of
``pathological'' type, that have to be studied in off-line simulations
\citep{sharpe1999a}.

The fundamental importance of hydrodynamical instabilities for the
propagation of deflagrations requires special modeling approaches. As
discussed in Sect.~\ref{sect:hydro_challenges}, the interaction of the
flame front with self-generated turbulence boosts the burning
efficiency.  Because only the largest scales of the turbulent cascade
are resolved, the effect of flame surface enlargement due to
interaction with turbulent eddies on smaller scales has to be
compensated by imposing an effective \emph{turbulent burning speed} on
the scale of numerical resolution. This effective turbulent flame
propagation velocity replaces the laminar flame speed in the level-set
equation. According to \citet{damkoehler1940a}, it scales with the
turbulent velocity fluctuations on the considered length
scale. Because of numerical dissipation, these are difficult to
determine close to the resolution of the computational grid, and
therefore turbulent subgrid-scale models are employed to determine
them (see
e.g. \citealp{niemeyer1995b,schmidt2006c,roepke2009a,ciaraldi2009a,hicks2013a,jackson2014a,roepke2017a}
for a discussion of approaches used in SN~Ia explosion models). It is
one of the important achievements of multidimensional simulations to
capture the effect of turbulent flame acceleration in a
self-consistent way.

Another challenge is the modeling of nuclear reactions that take place
in and behind the combustion wave. Two major obstacles have to be
overcome in this context. The first is that many reactions are
involved in the burning and an extended nuclear network is necessary
to predict the synthesis of all involved isotopes. Solving the full
network concurrently with the hydrodynamic simulation requires
substantial computational effort, in particular in three-dimensional
setups. While this is a practical challenge, the second is more
fundamental. If combustion waves are represented as discontinuities,
their internal structure and details of the reactions are not
captured. Artificially broadened combustion waves face the problem that
the length scales on which the species conversion and energy release
proceed physically are not resolved. They are also challenged by the
numerical effort of an extended nuclear network. For this reason,
reduced nuclear networks are usually employed in the hydrodynamic
explosion simulations, that follow only a few representative species
(accounting, for instance, for unburnt fuel material, intermediate
mass elements, and nuclear statistical equilibrium compositions). The
primary goal of the description of nuclear reactions in the hydrodynamic
explosion simulations is to model the energy release driving the
dynamics.  With reduced networks and artificially broadened combustion
waves, it is possible to approximate the energy release to a
sufficient accuracy. In models with very few representative species
and/or discontinuity descriptions of the combustion waves, the energy
release cannot be consistently reproduced and has to be
calibrated. This is either done on the basis of one-dimensional
resolved flame simulations or in an iterative procedure involving a
sequence of explosion models and nucleosynthesis post-processing
step. Such post-processing is also necessary to achieve the secondary
goal of modeling the burning processes: the determination of detailed
nucleosynthetic yields and the chemical structure of ejected material
in thermonuclear supernova explosions. The key idea is to place
virtual particles (so-called tracers) in the material of the exploding
white dwarf star so that each represents a certain fraction of the
total mass. These tracer particles are then passively advected with
the flow of the exploding material and record the thermodynamic
trajectories representative for the fraction of mass they follow. This
data is then used as input to a post-processing step that reconstructs
the details of the nuclear reactions based on an extended nuclear
reaction network \citep[see, e.g.,][]{travaglio2004a}.

The detailed hydrodynamic and chemical structure of the ejected
material is part of a modeling pipeline that follows the supernova
event from the progenitor structure over hydrodynamical explosion
simulations and nucleosynthetic post-processing to the formation of
observables that can then be compared to astronomical data. It is
input to multidimensional radiative transfer calculations that will be
discussed in Sect.~\ref{sect:obs}.

\subsection{Requirements for a viable explosion scenario}
\label{sect:requirements}

A fundamental goal of modeling thermonuclear explosions in white
dwarfs is to reproduce the characteristic spectral features of Type Ia
supernovae. The lack of hydrogen and helium is characteristic for this
class of objects. Moreover, spectral features indicate the presence of
substantial amounts of iron group and intermediate-mass elements. This
is prototypical for burning carbon-oxygen white dwarf
matter. Irrespective of the combustion wave being a deflagration or a
detonation, the released energy and the composition of the ash depends
on the fuel density ahead of the front.

At the highest densities, as encountered in the cores of massive white
dwarf stars, the ash temperatures become high enough to establish
nuclear statistical equilibrium (NSE) conditions. Freeze-out from NSE
occurs when the ejecta expand and iron group nuclei are formed. At
lower fuel densities, burning is incomplete and intermediate-mass
elements (Si, S, Ca, etc.) are synthesized. At even lower densities,
carbon burns to oxygen, and below a certain threshold, burning ceases
and unprocessed carbon-oxygen white dwarf material is left behind.

The fact that intermediate-mass elements are seen in the spectra
implies that a substantial amount of the stellar material must be
processed at sufficiently low densities ($\rho_\mathrm{fuel} \lesssim
10^7 \, \mathrm{g}\, \mathrm{cm}^{-3}$) to enable incomplete
burning. The burning front therefore must either (1) pre-expand a
Chandrasekhar-mass WD, which requires a sub-sonic flame propagation
mode, (2) proceed as a detonation in a pre-expanded Chandrasekhar-mass WD
in a delayed detonation scenario, or (3) form a detonation in a
sub-Chandrasekhar mass WD. We will discuss these possibilities in
Sect.~\ref{sect:simulations}.

\section{Prediction of observables}
\label{sect:obs}

\subsection{Radiative transfer considerations}

As outlined above, hydrodynamical explosion models can simulate the
dynamics and nucleosynthesis in thermonuclear supernovae from the
point of ignition until the ejecta reach near-homologous
expansion. However, to assess the validity of such models, the
explosion model output must be mapped onto the space of observable
quantities that can be compared to data. In general, this requires
additional calculations that yield predicted light curves, spectra
and/or spectropolarimetry. Fortunately, as noted above, such
calculations can usually be performed as a post-processing step on the
ejecta that have been dynamically simulated into the homologous phase.
There are important exceptions, however, most notably for scenarios
that involve ongoing dynamics as the explosion ejecta interact with a
dense environment \citep[see,
  e.g.,][]{fryer2010a,blinnikov2010a,noebauer2016a}.

Thermonuclear supernovae have ejecta that are rich in heavy elements:
in models of normal SNe~Ia, around half of the ejected mass is
composed of iron-peak elements, with around half of the remainder
being the so-called intermediate-mass elements (Si, S, Ca etc.). The
relative complexity of the atomic structure of the first few ions of
these elements (compared to e.g. H or He, which are more usually
dominant) means that large numbers of bound-bound transitions need to
be taken into account when simulating radiative transfer. Moreover,
the large expansion velocities blend the transitions together meaning
that it is usually very difficult to define a simple continuum or
separate out individual lines. As a consequence, the net contribution
of bound-bound transitions tends to dominate over any continuum
thermalization opacity \citep[see, e.g.][]{pinto2000b} and fluorescent
frequency redistribution becomes key to understanding the overall
spectral energy distribution.

The large expansion velocities (and associated velocity gradients)
have a central role in shaping the spectral features of supernovae and
need to be considered in any method that aims to predict synthetic
spectra. Several contemporary radiative transfer approaches embed the
assumption of high-velocity gradients in the form of the Sobolev
approximation \citep[see e.g.][]{sobolev1960a, lamers1999a}). This
approach makes it relatively easy to take into account very large
numbers of bound-bound transitions at modest computational cost
(either directly, e.g. \citealp{mazzali1993a}, or via an expansion
opacity formalism, e.g. \citealp{karp1977a,blinnikov1998a}). However, the
Sobolev approximation does have limitations, particularly in relation
to the treatment of overlapping lines \citep[e.g.][]{baron1996a}), which
becomes increasingly common at short wavelengths. Consequently, the
most sophisticated radiative transfer codes avoid this approximation
and treat individual line profiles in detail.

Reasonably accurate radiative transfer simulations also depend
strongly on the calculation of the temperatures and
ionization/excitation conditions in the ejecta. Local thermodynamic
equilibrium (LTE) is often adopted as a first estimate but departures
from equilibrium have important consequences and non-LTE effects
become increasingly important with time as the ejecta expand. As
illustrated by \citet{dessart2014b}, accurate synthetic observables
depend on describing a range of complicated microphysics, whose role
evolves with time. By the latest epochs commonly observed for
thermonuclear supernovae ($\sim 200 - 300$ days post explosion), the
ejecta are very far from LTE: at these "nebular" epochs, ionization
and heating controlled by the ongoing injection of non-thermal
particles (from radioactive decay) and cooling dominated by multiplets
of forbidden lines, predominantly of the iron group elements (for a
recent review of the modeling of nebular spectra, see
\citealp{jerkstrand2017a}). Consequently, modeling of spectra at these
epochs is critically dependent on the microphysics and
quality/quantity of the atomic data (radiative and collisional)
available for the necessary ions.

\subsection{Implementation and application to modern explosion models}

Given the competing demands on implementations and computational
resources (i.e. need to address complicated/time-dependent
microphysics in expanding, inhomogeneous 3D ejecta models), most
published studies to date have made several necessary
approximations. Currently, perhaps the most important trade-off made
in relation to the study of hydrodynamical explosion models is that
between simplified microphysics and multi-dimensional effects in the
ejecta. For example, while several of the Monte Carlo radiative
transfer codes (e.g. SEDONA, \citealp{kasen2006a}, or ARTIS,
\citealp{kromer2009a}) can compute orientation-dependent synthetic
observables for fully 3D ejecta models, these codes use the Sobolev
approximation and either use LTE or relatively simple nLTE
approximations. In contrast, sophisticated 1D codes can avoid the
Sobolev approximation and treat many levels of many ions in full nLTE
for SNe~Ia explosion models \citep[see, e.g.,][]{hoeflich1998a,
  baron2006a, blondin2013a}.

\section{Scenarios and simulations}
\label{sect:simulations}

\begin{figure}
\includegraphics[width=\textwidth]{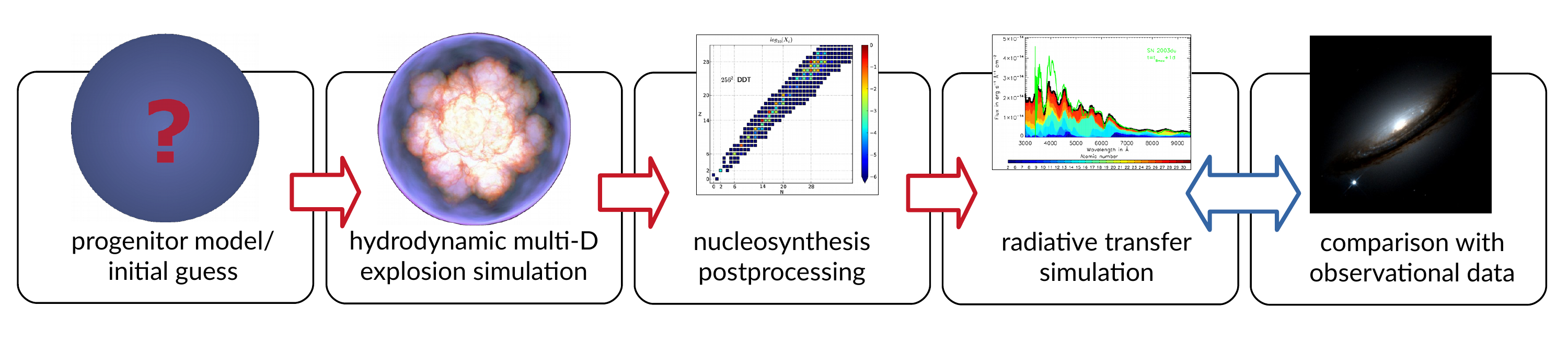}
\caption{3D modeling pipeline \label{fig:pipeline}}
\end{figure}

To meet the requirements for a viable explosion scenario discussed in
Sect.~\ref{sect:requirements}, different modes of ignition and flame
propagation are necessary, depending on whether the exploding star is
close to the Chandrasekhar-mass or below that limit. The first case
constitutes so-called Chandrasekhar-mass explosion models and the
second sub-Chandrasekhar mass models. We explore these model classes
below. They are addressed with our modeling pipeline shown in
Fig.~\ref{fig:pipeline}.

\subsection{Chandrasekhar-mass white dwarf explosion models}

Approaching the Chandrasekhar mass, the density in the core of a white
dwarf increases steadily. This will lead to the ignition of carbon
fusion in the so-called intermediate thermopycnonuclear regime
\citep{gasques2005a}, i.e.\ under conditions where the density has an
appreciable effect on the reaction rate. Initially, energy losses due
to neutrinos formed in plasmon decays and electron-nucleus
bremsstrahlung cool the stellar center. Ignition occurs when the
central density reaches high enough values so that neutrino losses become
insufficient to balance the energy production due to carbon burning.

This does, however, not yet trigger the explosion process. It rather
leads to a century of ``simmering'', in which convective motions
efficiently transport the energy generated in the stellar center
outward. The fluid motions are highly turbulent. On this background, a
hotspot finally develops, out of which a deflagration wave is formed
by thermonuclear runaway. Simulating the entire simmering phase is
virtually impossible, because a century cannot be bridged and the
spatial resolution is far from resolving turbulence. Nonetheless,
ignition simulations have been performed
\citep{hoeflich2002a,kuhlen2006a,zingale2009a,nonaka2012a}. A
three-dimensional simulation following the last hours until the first
thermonuclear runaway occurs at a radius of $\sim 50 \, \mathrm{km}$
off-center is presented by \citet{nonaka2012a}. The results indicate
that a second runaway at a different location shortly after the first
is unlikely. Thus, to current best knowledge, the deflagration will
form in a single region off-center of the WD star.

\subsubsection{Pure deflagrations}

\begin{figure}
\centerline{\includegraphics[width=0.65\textwidth]{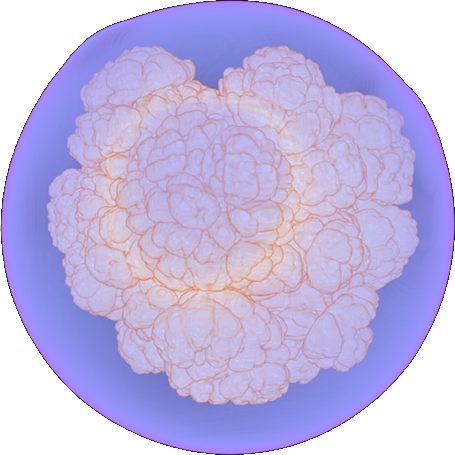}}
\caption{Simulation of a deflagration (orange/white contour) in a Chandrasekhar-mass white dwarf star (blue color). \label{fig:def}}
\end{figure}

The pure deflagration scenario follows possibility (1) described in
Sect.~\ref{sect:requirements}.  After formation of the deflagration
wave near the center, it propagates towards the surface, subject to
buoyancy instability. Multidimensional simulations clearly show the
formation of ``mushroom-shaped'' bubbles giving rise to a complex
morphology of the flame front (see Fig.~\ref{fig:def} for an
illustration). The flame strongly accelerates due to the interaction
with turbulence. The scaling behavior of turbulent motions in this
situation was unclear for a long time. Based on highly resolved
three-dimensional simulations, however, it could be shown that at
small length scales and for most of the burning turbulence is
isotropic and follows Kolmogorov scaling
\citep{zingale2005a,ciaraldi2009a}.

The strength of the deflagration and the overall outcome of the
explosion phase fundamentally depend on the initial
conditions. Several parameters are expected to vary in nature from
event to event, such as the central density and the chemical
composition of the exploding white dwarf star. Other parameters are
unknown or subject to uncertainties in the numerical modeling. A
parameter of models of the explosion phase is the ignition
geometry. The length scales of the actual flame formation cannot be
resolved in multidimensional simulations. Therefore, the effect is
usually mimicked by placing a number of flame kernels near the stellar
center. Although this does not necessarily capture the ignition
physics, it is a way of defining a well-posed initial setup. For a
single sphere, the Rayleigh-Taylor instability is seeded by random
numerical noise and imprinting a certain spectrum of resolved
perturbations on the flame geometry ensures convergence of the
model. The simulations of \citet{fink2014a} showed that the initial
flame shape has a tremendous impact on the strength of the burning. In
this study, a sequence of models was presented with varying numbers of
ignition sparks, that were randomly placed in the central region of
the white dwarf star. Sparse ignitions naturally lead to aspherical
flame geometry evolution, whereas an on average isotropic flame
propagation is only possible with dense ignitions. In the former case,
only a small fraction of the star is burned. Due to buoyancy, the
flame quickly rises towards the surface leaving the far side of the
star unaffected. The energy liberated in this process leads to the
ejection of parts of the stellar material, and a bound remnant is left
behind. In contrast, a complete disruption of the white dwarf star is
possible for dense isotropic ignition configurations. In all cases,
however, the ejecta structure is well-mixed on large scales due to the
flame instabilities. Even with dense ignition configurations, the
production of \isotope[56]{Ni} hardly exceeds $0.35 M_\odot$.

The relatively low $\isotope[56]{Ni}$ production means that pure
deflagration models generally fail to account for the observed
brightness of the majority of SNe~Ia (the predicted peak luminosity of
such models is too low). However, the $\isotope[56]{Ni}$ masses are
consistent with those required for some low-luminosity events, of
which several sub-classes have now been identified
\citep{taubenberger2017a}. In particular, the range of
$\isotope[56]{Ni}$ masses predicted in pure deflagration models
\citep{jordan2012b,fink2014a} is roughly consistent with the observed
range of brightness spanned by the Type~Iax supernovae
\citep{foley2013b}.

The potential identification of Type~Iax supernovae with pure
deflagrations is supported by comparison of synthetic spectra and
light curves to observations. For example, \citet{kromer2013a} compared
model predictions for one of the pure deflagration models of
\citet{fink2014a} for a range of photospheric-phase epochs to SN2005hk
\citep{phillips2007a}, a well-observed member of the SNe~Iax
class. They found fair agreement in both the strengths and shapes of
spectral features across a range of phases in the optical and
infrared, and also good correspondence between the model and observed
colors around maximum lights. However, some important discrepancies do
remain. In particular, the model light curves evolve too quickly, most
notably in the post-maximum decline phase in the redder bands
\citep{kromer2013a}. This systematic discrepancy is also apparent in
the comparison of a different model from the \citet{fink2014a} sample
to a fainter member of the SNe~Ia class, 2015H
\citep{magee2016a}. Several studies have also drawn attention to
evidence that the ejecta of SNe~Iax are not fully mixed
\citep{stritzinger2015a,barna2017a}, which is difficult to reconcile
with a turbulent deflagration model.
In addition, it remains unclear whether pure deflagration models can
account for the lowest luminosity members of the SNe~Iax class, such
as SN2008ha \citep{foley2009a}, which requires less than $0.01 M_\odot$
of \isotope[56]{Ni}. Such a low mass of \isotope[56]{Ni} might be
achieved under conditions whereby burning in the deflagration is
occurs only in a limited central region of the WD, for example due to
an exotic composition \citep{kromer2015a} -- however, it remains to be
demonstrated whether this can be realized in nature.

One outstanding, but noteworthy, feature of the comparison of SNe~Iax
and pure deflagration models is potential role of the residual
material from the WD, that was still bound at the end of the explosion
phase (in e.g. the models compared to observations mentioned above,
$\sim$$1 M_\odot$ or more of the mass of the initial WD remains bound
at the end of the explosion simulation). Some of the
$\isotope[56]{Ni}$ synthesized in the explosion remains in this
material \citep{kromer2013a,fink2014a}, meaning that it will experience
ongoing energy injection which will plausible drive further expulsion
of mass \citep{foley2016a}. Further study, both of physical conditions
in the residual material \citep{shen2017a} and of late-phase
observations of SNe~Iax \citep{foley2016a} are needed to explore this
topic in more detail.

\subsubsection{Delayed detonations}

Enhancing the \isotope[56]{Ni} production and the explosion energy to
values necessary to reproduce normal SNe~Ia is not possible by simply
increasing the number of ignition kernels or tuning the initial
parameters of the exploding white dwarf. A fundamental change in the
burning mode is required -- a transition from the initial
deflagration, that is necessary to pre-expand the material, to a
subsequent detonation. This scenario follows possibility (2) described
in Sect.~\ref{sect:requirements} This is the idea of the class of
\emph{delayed detonation models}. The key question in these is
obviously if and how a detonation is triggered in a late burning
stage.

Several mechanisms have been proposed for initiating detonations in
delayed detonation models. A spontaneous deflagration-to-detonation
transition (DDT) may be caused by intrinsic processes in the burning
wave.  Whether or not such DDTs occur in thermonuclear supernova
explosions remains uncertain. Some necessary conditions were laid out
in the studies of
\citep{lisewski2000a,woosley2007a,woosley2009a,woosley2011a}.  Two
other mechanisms, the gravitationally confined detonation (GCD,
\citealp{plewa2004a}) and the pulsational delayed detonation (PDD)
mechanisms \citep[e.g.,][]{bravo2006a}, rely on weak initial
deflagration stages that fail to gravitationally unbind the white
dwarf star.

The mechanism for GCD assumes that a one-sided ignition leads to an
asymmetric deflagration that is too weak to gravitationally unbind the
WD star. The deflagration ash breaks out of the star's surface and
sweeps around it to collide in the antipode. Clearly, a successful
ignition of a detonation in this collision favors stronger impact
which in turn implies a weak deflagration phase. The resulting
detonation then burns the bound core of the object. With weak
deflagrations, it will not be very expanded and thus particularly
bright events with high masses of synthesized $^{56}$Ni are
expected. A bound white dwarf resulting from a weak deflagration will
pulsate. These pulsations may aid the formation of a detonation
\citep{jordan2012a}. Both the GCD and the PDD scenarios share the
characteristic feature that the products of high-density deflagration
will be located in the outer part of the ejected material at high
velocities. This is in conflict with observations
\citep{seitenzahl2016a}.

Also for the classical (DDT) delayed detonation scenario, several
problems persist. It has been suggested as a model for the bulk of
normal SNe~Ia. This requires them to reproduce individual supernova
observations. Studies based on 1D DDT models have generally been
fairly successful in this regard \citep[e.g.][]{hoeflich1998a,
  blondin2013a}. Radiative transfer simulations based on
multi-dimensional simulations of DDTs (e.g., in 2D \citealp{kasen2009a},
\citealp{blondin2011a}, or in 3D \citealp{seitenzahl2013a}, \citealp{sim2013a})
have also generally found that some DDT models can provide a fairly
good (albeit far from perfect) match to many properties of the light
curves spectra, and indeed spectropolarimetry \citep{bulla2016b} of
individual SNe~Ia.

In addition, if DDT models are responsible for the full population,
observed trends between characteristic features should be
reproduced. The model explosions should be able to cover the range of
brightnesses observed from normal SNe~Ia. This requires a nickel mass
production in the range from below $~0.4 \, M_\odot$ to $0.8 \,
M_\odot$, see e.g. \citet{scalzo2014a}. Delayed detonations face a
fundamental challenge here. Generally, stronger deflagrations lead to
increased expansion before the detonation phase sets in
\citep{roepke2007b,mazzali2007a}. Consequently, the detonation runs
over lower-density material and produces less $^{56}$Ni. Therefore,
the faintest models are expected to be those with the strongest
deflagration. This was tested in multi-spot ignition setups that allow
to vary the deflagration strength significantly. The strongest
deflagrations produce $\gtrsim 0.3 \, M_\odot$ of $^{56}$Ni and in the
subsequent detonation little is added to this amount. This means that
multi-spot ignitions with many, on average isotropically distributed
kernels are required to reach the fainter end of the distribution of
normal SNe~Ia. These, however seem unlikely to be realized in nature
\citep{nonaka2012a}. Furthermore, when such models are invoked, they
appear to fail to fully produce the relatively rapid light curve
evolution that is observed to coincide with low luminosity (i.e. the
light curve width-luminosity relation: see e.g. \citet{sim2013a}).  The
second problem in this context is that the brightness distribution of
normal events is observed to peak at explosions producing $\sim 0.6 \,
M_\odot$ of $^{56}$Ni. It is not obvious why the initial parameters
such as flame ignition geometry, central density and chemical
composition of the progenitor white dwarf star, or the initiation
mechanism of the detonation, should favor a configuration producing
this amount of radioactive nickel.

We note that, although the 3D simulations of \citet{sim2013a} have
difficulties reproducing the observed width-luminosity relation with a
faster decline of the $B$-band light curve for fainter events, it may
be possible to construct such a relation in Chandrasekhar-mass white
dwarf star explosions \citep{kasen2007a}. Recent studies
\citep{blondin2017a,goldstein2018a}, however, increasingly indicate
that sub-Chandrasekhar mass explosions are required to capture the
full range of the observed relation.

\subsection{Sub-Chandrasekhar mass white dwarf explosions}

The alternative to Chandrasekhar-mass white dwarf explosions as models
for SNe~Ia are detonations in sub-Chandrasekhar mass objects following
possibility (3) of Sect.~\ref{sect:requirements}. The capability of
the scenario to reproduce basic characteristics of observed SNe~Ia was
demonstrated by sequences of toy models. Here, if white dwarfs of
varying mass are set up and a central detonation ignited artificially,
the resulting ejecta structures provide a relatively simple sequence
of models that do a fair job of reproducing many of the observed
characteristics of SNe~Ia \citep[e.g.][]{shigeyama1992a,
sim2010a, shen2017a}. A very attractive feature of this
class of model is that a single simple physical parameter can be
recognized as responsible for driving many of the differences between
different explosions: the mass of the exploding WD. In particular, a
wide range of nickel mass can be produced from only a modest range of
WD masses: an initial WD mass of around 1 M$_{\odot}$ is required to
produce an explosion with brightness characteristic of the most common
SNe~Ia \citep[see][]{sim2010a,shen2017a}. In addition, the low densities
in sub-Chandrasekhar mass white dwarfs means that their detonation
yields significant quantities of intermediate mass elements, allowing
for a relatively good match to observed spectra across a broad range
of explosion luminosity. One important challenge for this class of
simple (toy) sub-Chandrasekhar mass models is that they struggle to
sustain sufficiently slow light curve evolution to account for the
brighter end of the SNe~Ia distribution. However, the variation in
ejecta mass amongst sub-Chandrasekhar mass models does naturally
suggest a link between brightness and light curve evolution and
simulations have favored sub-Chandrasekhar models for relatively faint
explosions \citep[see, e.g.,][]{blondin2017a}.

The problem with sub-Chandrasekhar mass explosion models of the sort
described above is that the physical ignition of detonations in such
objects does not intrinsically arise from an evolutionary process
(such as the accretion of mass towards the Chandrasekhar limit).  It
has to be caused by some vigorous event. Two possibilities are
commonly discussed.

One way to ignite a detonation in a carbon-oxygen white dwarf star is
that it accretes helium-rich material from a companion. Due to
instabilities in the accretion process or once the accreted shell has
grown massive enough, a detonation triggers in the He material. It
propagates the carbon-oxygen core and drives a shock wave into
it. This shock wave may trigger a secondary detonation in
carbon-oxygen rich material -- constituting the so-called \emph{double
  detonation explosion scenario} (see \citealp{fink2010a,moll2013a} for
recent multidimensional supernova simulations following this
paradigm). It is conceivable that this occurs when the shock hits the
outer edge of the core (``edge-lit double detonation'') or due to a
spherical collimation of the inwards propagating oblique shock wave
near its center. The latter case was shown to robustly lead to
detonations of the core by a geometric amplification effect
\citep{fink2007a}. It has to be emphasized, however, that many of the
published models simply \emph{assume} a primary detonation of the He
shell. The ignition process is very hard to resolve numerically and
the success of the scenario hinges on it to occur in reality.
Synthetic spectra have been computed from double detonation models for
a range of conditions \citep[e.g.][]{nugent1997a, kromer2010a,
woosley2011b}, with results that depend significantly on the
assumed structure and mass of the helium layer at explosion. Indeed,
it generally appears to be the case that the influence of the outer
ejecta layers (rich in helium and/or helium-detonation ash) is mostly
detrimental to the agreement of the models with observations: if
substantial helium shells are invoked this leads to effectively
suppressing the characteristic features of intermediate mass elements
and to very dramatic line blanketing effects associated with heavy
elements that are synthesized in the helium detonation. Thus, for such
double detonation models to be viable, very low mass helium shells
must be invoked: whether such low masses of helium can really be
ignited and/or sustain a detonation is a topic of active study
\citep[e.g.][]{shen2007a,shen2007a,shen2014a}.

An alternative scenario is that of \emph{violent mergers}
\citep{pakmor2010a,pakmor2011b,pakmor2012a}. In contrast to the
classical merger paradigm, the explosion happens before the two white
dwarfs have formed a single object. In the inspiral process, the
lighter of the pair is disrupted and its material plunges into the
more massive primary that is only weakly affected by tidal
forces. This impact may trigger a detonation of the primary -- a
sub-Chandrasekhar mass white dwarf. A recent update of the model
\citep{pakmor2013a} suggests that the ignition of the detonation is
triggered even earlier in the inspiral process when He rich material
(that always exists in low quantities on top of a carbon-oxygen white
dwarf) is accreted from the secondary to the primary. This rapid
accretion process leads to hydrodynamical instabilities in the
He-layer on the primary and triggers a detonation in this
shell. Similar to the double detonation scenario, the actual
supernovae results from a secondary detonation of the core
material. In contrast to that scenario, however, in the violent merger
case the He shell is less massive and less dense so that its imprint
on the predicted observables is much reduced.

It is remarkable that population synthesis studies predict a peak of
the distribution of white dwarf mergers at primary masses that produce
$\sim$$0.6 \, M_\odot$ of $^{56}$Ni \citep{ruiter2013a}. Moreover, also
the temporal evolution of the luminosity function resulting from
sub-Chandrasekhar mass WD detonations seems to match the observations
\citep{shen2017a}.

\section{Conclusions}

Thanks to advances in computing resources and numerical methods in
recent years, we are now able to perform meaningful fully 3D explosion
simulations for a range of progenitor scenarios that have been
proposed for Type Ia supernovae. Combined with radiative transfer
post-processing, which allows predictions to be made that can be directly
compared to observations, such simulations are now playing a key role
in driving our understanding of the nature and physics of
thermonuclear supernovae. However, the state of the art remains
incomplete and far from satisfactory -- numerous limitations
persist. These include clearly posing initial conditions for explosion
simulations in the context of particular progenitor modeling, proper
representation of the dynamics and instabilities during the
thermonuclear combustion in full star models, and adequate description
of the complex radiation processes responsible for spectrum formation
in the evolving ejecta. The last decade has demonstrated that such
multi-dimensional simulations are possible. The goal for the future
will be their development towards a level of predictive power than
allows for ever-improving quantitative testing by comparison to the
increasing wealth of observational data.

\begin{acknowledgements}
The work of FKR is supported by the Klaus Tschira Foundation and by
the Collaborative Research Center SFB 881 ``The Milky Way System''
(subproject A10) of the German Research Foundation (DFG).
\end{acknowledgements}


\end{document}